\documentclass[useAMS,usenatbib,a4paper, fleqn]{mnras}
\usepackage{ae,aecompl}
\title[Evolution of spiral structure]
{The evolution of pitch angles of spiral arms
}
\author[Pringle]
{J. E. Pringle\thanks{E-mail:
jep@ast.cam.ac.uk}$^{1}$, C. L. Dobbs$^{2}$ \\
$^1$ Institute of Astronomy, Madingley Road, Cambridge, CB3 0HA \\
$^2$ School of Physics and Astronomy, University of Exeter, Stocker Road, Exeter, EX4 4QL, UK \\
}

\usepackage{amssymb}
\usepackage{amsmath}
\usepackage[pdftex]{graphicx}
\usepackage{epsfig}
\usepackage{multirow} 
\def\aap{{ A\&A}}

\def\apjl{{ApJL}}

\def\apjs{{ApJS}}

\begin{document}
\label{firstpage}
\date{\today}

\pagerange{\pageref{firstpage}--\pageref{lastpage}} \pubyear{2019}

\maketitle

\begin{abstract}

In spiral galaxies, the pitch angle, $\alpha$, of the spiral arms is often proposed as a discriminator between theories for the formation of the spiral structure. In Lin-Shu density wave theory, $\alpha$ stays constant in time, being simply a property of the underlying galaxy. In other theories (e.g tidal interaction, self-gravity) it is expected that the arms wind up in time, so that to a first approximation $\cot \alpha \propto t$. For these theories, it would be expected that a sample of galaxies observed at random times should show a uniform distribution of $\cot \alpha$. We show that a recent set of measurements of spiral pitch angles (Yu \& Ho 2018) is broadly consistent with this expectation.

\end{abstract} 


\section{Introduction}
Spiral galaxies make up around one third
of all massive galaxies \citep{Lintott2008}. Star formation in the current Universe overwhelmingly occurs in spiral galaxies, and in particular in the spiral arms. How the spiral arms form and evolve (if they do) is still a matter for debate. The properties of spiral galaxies, as well as the origins and the nature of the spiral structure itself, are reviewed by \citet{DobbsB2014}. They argue that (at least in unbarred spirals) spiral arms are either transient, recurrent entities, being the result of self-gravity within the stellar and/or gaseous disc, or are the result of tidal interactions. This picture is supported by recent observations from Galaxy Zoo results showing that there is no link between spiral arms pitch angle and central galaxy concentration, which is inconsistent with a modal theory of spiral arms \citep{Hart2017}. \citet{Shabani2018} also analyse age gradients in 3 galaxies and only find evidence of a fixed density wave in a galaxy with a strong bar. In contrast, after analysing the properties of a large sample of spiral galaxies, \citet{Yu2018,Yu2019} (see also \citealt{Yuall2018}) do find correlations of pitch angle with galaxy morphology, which lead them to favour the idea that spiral arms are manifestations of standing modes of propagating spiral waves within the galaxy \citep{Lin1964}.

In Section~\ref{generation} we summarise briefly the properties of the various mechanisms for the generation of spiral arms, with emphasis on their different predictions for the evolution of the spiral patterns. 
In particular, we use the pitch angle, a key diagnostic in the aforementioned observational work, as a measure of the shape of the spiral arms and predict the evolution of this characteristic over time.
We draw attention to the fact that only the modal density wave theory predicts that the spiral pitch angles are constant in time. Of course, for an individual galaxy it is not possible to observe any change of pitch angle with time. However, in Section~\ref{test} we show that the possibility of pitch angle evolution does give rise to predictions of what the distribution of pitch angles should be for a randomly chosen ensemble of galaxies. We propose a simple test of this hypothesis and show that these predictions are  amply satisfied by the data. In Section~\ref{discuss} we conclude that this finding is fully consistent with the ideas of spiral arms being generated by either internal self-gravity or tidal interactions \citep{DobbsB2014}.

\section{Generation of spiral structure}
\label{generation}

Here we describe briefly the leading ideas for the different mechanisms of spiral arm generation in non-barred galaxies. And, in addition, we draw attention to what the different mechanisms predict for spiral arm evolution.

\subsection{Density wave theory -- Stationary spiral structure modes -- Lin-Shu hypothesis}
\label{LS}

Density perturbations with spiral spatial structure can propagate radially within a self-gravitating disc of gas or stars. The Lin-Shu hypothesis \citep{Lin1964} proposes that spiral structure consists of a stationary density wave, which is basically a standing mode. Thus the spiral pattern remains unchanged, except for an overall rotation at some fixed radius-independent, pattern speed
\begin{equation}
\label{LSEQ}
 \Omega_p = {\rm const.},
 \end{equation}
 for many galactic rotation periods. These can be thought of as global modes of the galaxy as a whole, and their properties therefore must depend on global properties of the galaxy. Recent attempts to find evidence for this can be found in \citet{Grand2013} and  \citet{Yu2019}.

\subsection{Tidal interactions}
\label{TI}

The idea that tidal interactions are a dominant reason for the generation of spiral arms became firmly established following work by \citet{Toomre1972}. Typically, the dominant gravitational perturbation in a tidal interaction is of the form $m=2$, where $m$ is the azimuthal wavenumber. In a galactic disc which is not subject to self-gravity, the most long-lasting response to such a perturbation is a kinematic density perturbation\footnote{This is often referred to as a kinematic density wave, but it is not really a wave in the sense that, in the absence of self-gravity, it does not propagate radially within the disc.}. Such a perturbation evolves into a spiral pattern with the radially dependent pattern speed
\begin{equation}
\label{TIEQ}
\Omega_p(R) = \Omega(R) - \frac{1}{2} \kappa(R),
\end{equation}
where $\Omega$ is the galactic rotation rate and $\kappa$ the local epicyclic frequency.\footnote{It should be noted, that because both the kinematic disturbances discussed here, and the modal Lin-Shu density waves discussed in Section~\ref{LS}, are such that for most (and often all) disc radii $\Omega_p \neq \Omega(R)$, the observation of a wavelength dependence of spiral pitch angle is not a discriminator between them (cf. \citealt{Dobbs2010b, Yu2018,  Miller2019}).}

Weak self-gravity within the disc provides small modifications to this result, in particular the possibility of radial wave propagation. And it is worth noting that galaxy-galaxy interactions are often more complicated than a single fly-by. For example, the spiral pattern in M51 appears to have been caused by the double interaction with the perturbing galaxy NGC5194 \citep{Dobbs2010}, and the complicated spiral structure of M81 appears to be the result of recent tidal interactions with the two galaxies M82 and NGC3077 \citep{Yun1999}.

\subsection{Recurrent, transient spiral instabilities, driven by self-gravity of the disc}
\label{SG}

If the disc of a galaxy is sufficiently self-gravitating (whether in the stars or the gas) local self-gravitational instability manifests itself in the form of local spiral density enhancements being sheared by the local galactic rotation. N-body simulations demonstrate that if the instability is sufficiently widespread these structures, although transient, can recurrently reform. In some simulations the arms appear to undergo a cycle, breaking into smaller segments and then reconnecting, due to differential rotation, to reform large-scale spiral patterns, whereas in others, where disc self-gravity is more influential, the large-scale arm patterns last for many rotation periods \citep{Fujii2011, Wada2011, D'Onghia2013, Baba2013}. 

This implies that the spiral arms locally co-rotate approximately with the galactic rotation, and therefore that these arms have an approximate pattern speed
\begin{equation}
\label{SGEQ}
\Omega_p(R) \approx \Omega(R).
\end{equation}

\section{A simple-minded test}
\label{test}

Following \citet{Binney2008} (Section 6.1) we define the pitch angle $\alpha$ of a spiral arm at any radius $R$ to be the angle $\alpha$ between the tangent to the arm and the circle radius $R = $ const. Then the pitch angle can be written as
\begin{equation}
\cot \alpha = R t \:  \bigg| \frac{d \Phi}{d R} \bigg|
\end{equation}
where $\Phi(R,t)=\Phi_0(R) +\Omega_p(R)  \: t$ is the equation describing the evolution of the azimuthal phase of a spiral arm with pattern speed $\Omega_p(R)$. By evaluating $\cot \alpha$ at times $t$ and $t_0$, we then obtain an equation for the evolution of $\cot \alpha$  with time, 
\begin{equation}
\label{alphavt}
\cot \alpha = \bigg[R \frac{d \Omega_p}{dR}\bigg] \: (t - t_0)+ \cot \alpha_0,
\end{equation}
where $\alpha = \alpha_0$ at time $t = t_0$. Thus $\cot \alpha$ evolves linearly in time.

The implication is that in the modal Lin-Shu picture (Section \ref{LS}), $\alpha$ remains constant in time, whereas, in both the tidal (Section \ref{TI}) and the self-gravitating (Section \ref{SG}) pictures, $\alpha$ decreases monotonically with time. Simulations of both tidal interactions (\citealt{PettittW2018}, Fig 19) and of transient, recurrent arms (\citealt{Grand2013}, Fig. 7) show the validity of  this expectation.

Given this, and given some simple (possibly over-simple) assumptions, we propose a test of Equation~\ref{alphavt}. We suppose that each galaxy in a sample starts to show measurable spiral structure at some maximum pitch angle $\alpha_{\rm max}$. This pitch angle then evolves for that galaxy, according to Equation~\ref{alphavt} until some minimum pitch angle, $\alpha_{\rm min}$ at which spiral structure is no longer apparent or measurable. We then also suppose that the pitch angle for spiral features within a galaxy is not strongly dependent on radius, so that the radius which gives the dominant contribution to a measure of the pitch angle does not change greatly with time. If these assumptions are valid, then  for each galaxy, $\cot (\alpha)$ evolves uniformly in time from $\cot (\alpha_{\rm max})$ to $\cot (\alpha_{\rm min})$. And, although the rate of evolution is different for different galaxies, provided that galaxies are observed at random times within this evolution, the distribution of $\cot \alpha$ for each galaxy, and therefore also for an ensemble of galaxies, should be uniform. To summarise, if we observe a random collection of galaxies at any particular time  we expect to find equal numbers in each equal-sized bin in $\cot \alpha$-space.

\citet{Yu2018} have measured the pitch angles of the spiral structure 113 galaxies selected from the Carnegie-Irvine Galaxy Survey. Pitch angles are measured at a range of wavelengths and using two different methods. The results show strong consistency between the methods, and strong correlations between different wavelengths. We consider the pitch angles measured by their 1D method, firstly in the visual band (\citealt{Yu2018}, Table 1, Column 19). For these 95 galaxies we show in Figure 1 the distribution in $\cot \alpha$. We have not attempted to make use of the error estimates provided for the values of $\alpha$ by \citet{Yu2018}, nor have we made any effort to make allowance for any systematic effects, such as the dependence on spiral measurement accuracy as a function of galactic inclination (See Section~\ref{discuss}). Nevertheless, Figure 1 does at least seem to be consistent with the idea that for this set of galaxies $\cot \alpha$ is more or less uniformly distributed within the range $1.00 \le \cot \alpha  \le 4.75$, that is from $\alpha \approx 45^\circ$ to $\alpha \approx 12^\circ$.

\begin{figure}
\centerline{\includegraphics[scale=0.42]{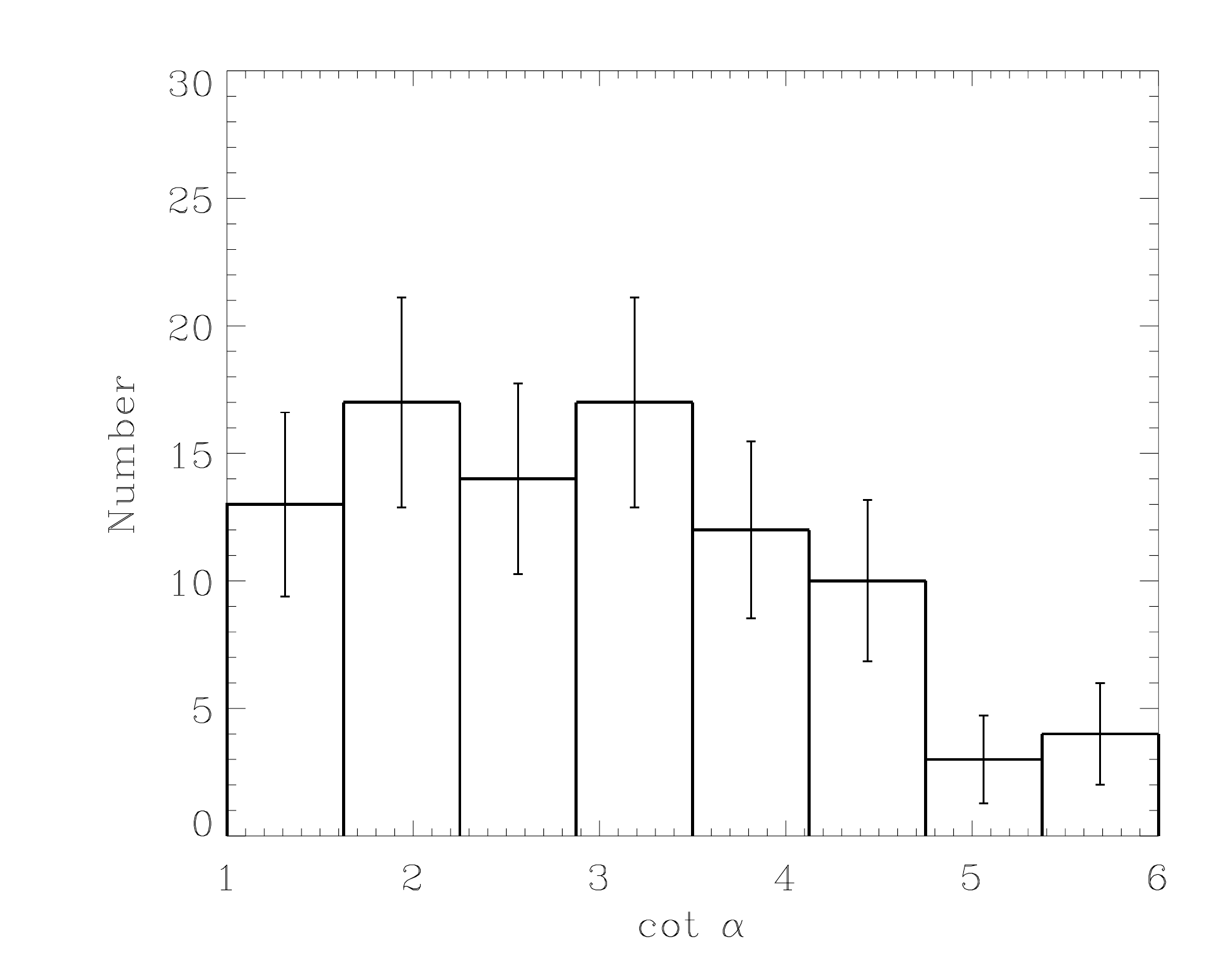}}
\caption{The distribution of pitch angles is shown using the data from \citet{Yu2018}.}
\end{figure}

To provide some statistical basis for this result we consider the 86 galaxies with $1.19 \le \cot \alpha \le 4.91$.  We plot in Figure 2 the cumulative distribution of these galaxies in $\cot \alpha$-space and compare it with the cumulative probability distribution function for a uniform distribution (i.e. a straight line). The Kologorov-Smirnov statistic for these distributions is $D = 0.128$ indicating consistency. 

\begin{figure}
\centerline{\includegraphics[scale=0.42, bb=0 200 600 650]{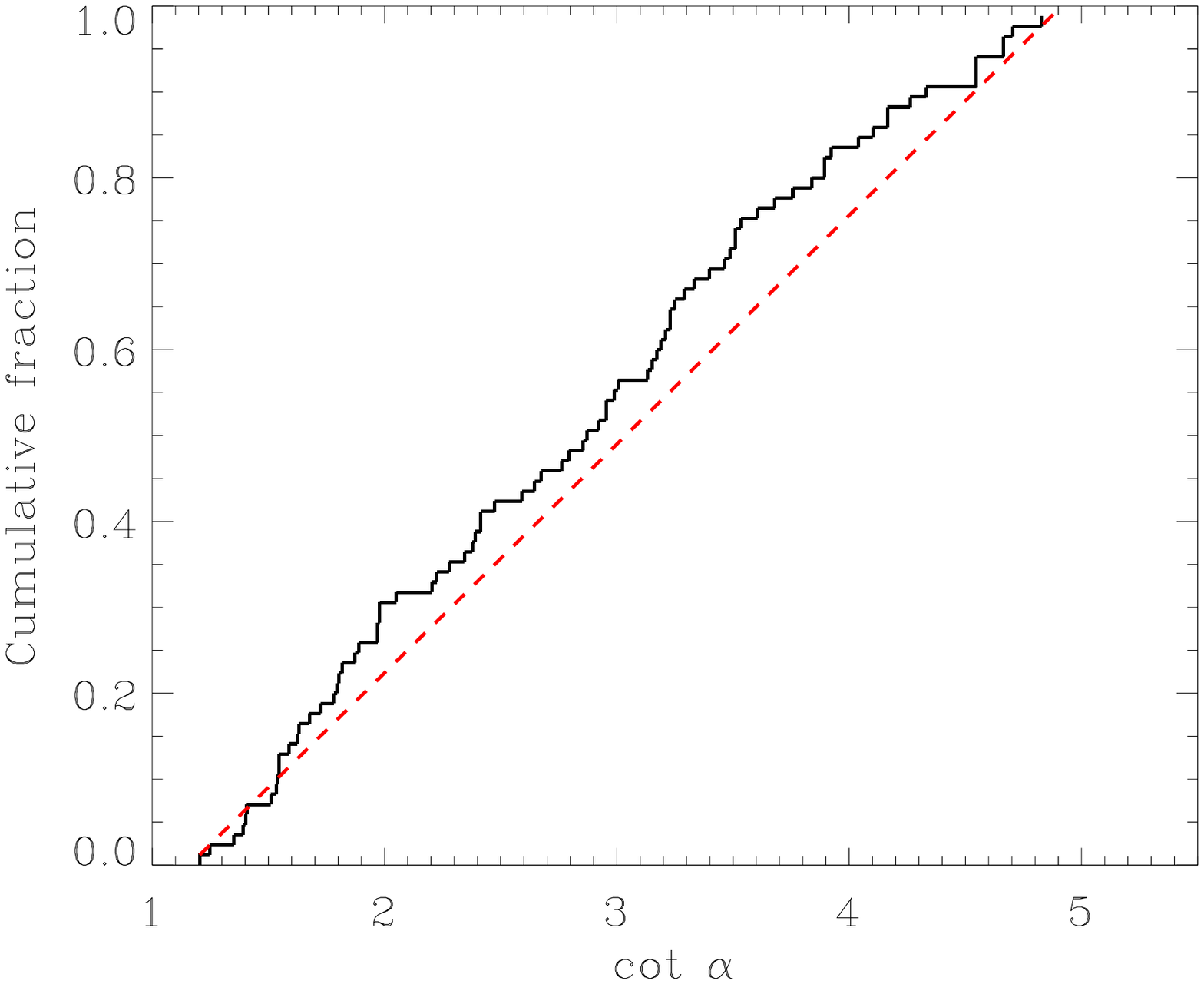}}
\caption{The cumulative fraction of $\cot \alpha$ where $\alpha$ is the pitch angle is shown, using the central 90\% range of pitch angles from \citet{Yu2018}. The dashed line shows a uniform distribution which is what would be expected if the spiral structure winds up at a constant rate (Equation (\ref{alphavt})). The KS statistic for the comparison between these two distributions is $D = 0.128$, confirms a lack of statistical disagreement.}
\end{figure}

As a consistency check, we also show the normalised distribution of pitch angles for different wavebands in Figure~3, again using data from \citet{Yu2018}. As would be expected, the distribution of pitch angles are similar for the different wavebands. Within error bars (which we do not show for clarity), the pitch angles again exhibit a uniform distribution in the range $1.00 \le \cot \alpha  \le 4.75$ for the B, V, R and 3.6$\mu$m bands. We leave any further studies across different wavelengths or other parameters to future observational studies.

\begin{figure}
\centerline{\includegraphics[scale=0.4, bb=0 200 600 650]{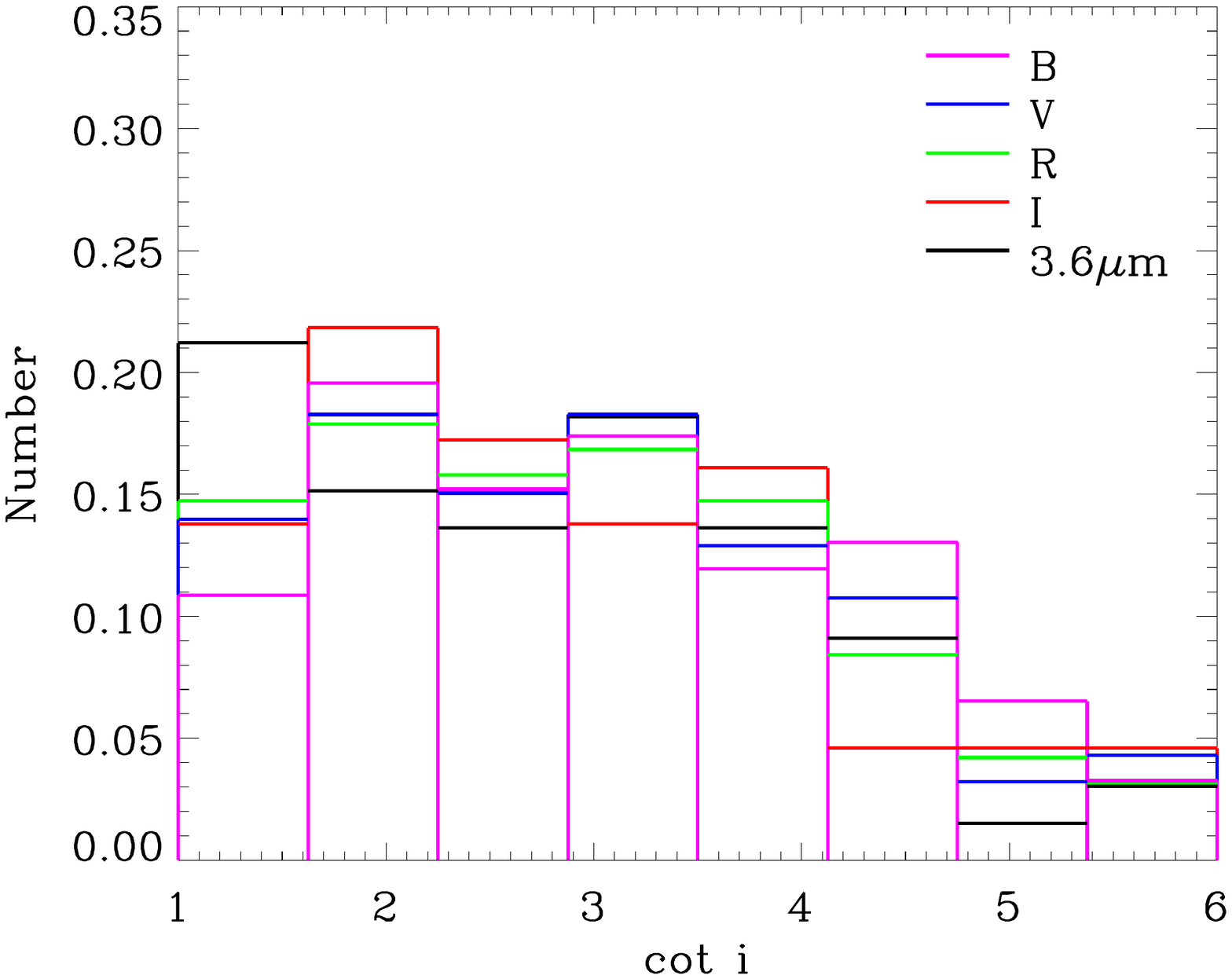}}
\caption{The distribution of pitch angles is shown for different wavebands, using the data from \citet{Yu2018}.}
\end{figure}

\section{Discussion}
\label{discuss}
Recent work has attempted to correlate measurements of pitch angles with other galaxy properties as a means to distinguish between mechanisms for spiral arm generation. In these observational studies the pitch angle is usually assumed to have a fixed value independent of radius or time. In particular Yu, Ho and others \citep{Yuall2018, Yu2018, Yu2019} argue that measurements of pitch angles, and the correlations between these and other galaxy properties, are broadly consistent with the expectations of Lin-Shu modal density wave theory (Section~\ref{LS})\footnote{although the opposite conclusion is reached from the analysis by \citet{Hart2017}.}.  We find nothing that contradicts this possibility. However, their arguments and analysis are predicated on the basic assumption of Lin-Shu theory, which is that each galaxy contains a spiral pattern  whose pattern speed, $\Omega_p$, and pitch angle, $\alpha$, are fixed in time (and space), being just dependent on the various properties of the galaxy, for example on morphology of the galactic rotation curve (Figure 11 in \citealt{Yu2019}) or the galactic shear rate (Figure 12 in \citealt{Yu2019}). Here we investigate whether an alternative premise, i.e. that the pitch angles of galaxies evolve with time, is consistent with the observed data.

Our finding that the distribution of pitch angles ($\cot \alpha$) is more or less uniform among their observed galaxies, opens up the possibility that the pitch angles are actually evolving in time, as is found in the simulations (e.g. \citealt{Grand2013, PettittW2018, Mata-Chavez2019}). 
We note that in the modal density wave picture, this finding would need some other explanation. We argue here that the possibility of evolving pitch angles merits further investigation.

Our simple assumption that the distribution of $\cot \alpha$ is uniformly distributed between  $\alpha_{\rm max}$ and  $\alpha_{\rm min}$ is strictly valid only if all galaxies have the same values of  $\alpha_{\rm max}$ and  $\alpha_{\rm min}$ and the evolution in $\cot \alpha$ proceeds in each galaxy at a constant (possibly galaxy dependent) rate, as in Equation (\ref{alphavt}). Even if the rate of evolution of $\cot \alpha$ were constant, a spread in the values of  $\alpha_{\rm max}$ and  $\alpha_{\rm min}$, would imply drop-offs in the $\cot \alpha$ distribution at each end.

Thus we need to consider the following:
\begin{enumerate}

\item What determines $\alpha_{\rm max}$? There is presumably some spread in the actual values of  $\alpha_{\rm max}$ as spiral structure is initiated. The largest values of $\alpha$ found by \citet{Yu2019} are around $\alpha \approx 40^\circ$. In the tidal case, the values of  $\alpha_{\rm max}$ are  likely to be determined by the strength of the tidal interaction, and perhaps by selection effects which exclude strongly interacting tidal pairs from the surveys. For the transient spiral case, the largest values of  $\alpha_{\rm max}$ are likely to result from the properties of individual galaxies (see, for example, \citealt{Grand2013, Mata-Chavez2019}).

\item What determines $\alpha_{\rm min}$? For small  $\alpha_{\rm min}$, it seem likely that the deciding factor is the observability of spiral features with small values of $\alpha$. Such features would be harder to detect in the more edge-on galaxies. For the tidal case, as time progresses the strength of the induced spiral feature is likely to decay, and this too can set a lower limit to the observability of  $\alpha$. For the transient spiral case, it is apparent that as the arms wind up, they may break up and reform into new global spiral structures (see, for example, \citealt{Baba2013, Mata-Chavez2019}). This process could also set a lower limit to $\alpha$, possibly dependent on galactic properties.

\item How constant in time is the evolution of $\cot \alpha$? 

The assumption underlying the analysis is that the pitch angle of a particular observed galaxy is independent of radius, and therefore that the pattern speed is independent of radius.
In the instances where galaxies do not exhibit steady state density waves, we would not expect this assumption to hold exactly. Observationally, previous studies have indicated a variation in pitch angle along spiral arms \citep{Savchenko2013, Honig2015}. There is mixed evidence regarding the pattern speed of spiral arms, \citet{Peterkin2019} claim that UGC 3825 exhibits fixed spiral arms, whereas other work has found a variation of pattern speed with radius \citep{Meidt2008, Speights2011,  Speights2012}. The methods used, for example by Hart etl al. (2017),   \citet{Yuall2018} and others, are likely to measure the pitch angle in a manner weighted to the radius at which the spiral structure is most apparent (as is found, for example, in the simulations of \citealt{Mata-Chavez2019}). If that radius were to change in time, then the rate of change of $\cot \alpha$ would not be strictly constant. The degree and manner of such a lack of constancy, would depend, among other things, on the morphology of the individual galaxy's rotation curve. 

\end{enumerate}

In summary, we have demonstrated that the distribution of spiral pitch angles in the galaxies investigated by \citet{Yuall2018} is consistent with the idea that the pitch angles evolve in time in the simple manner indicated in Equation (\ref{alphavt}). This indicates that the idea that most spiral structure is generated by tidal interactions and/or by internal self-gravity is still viable. We have suggested ways in which numerical simulations of these formation mechanisms could be used to shed light on the pitch angle distribution and its observed correlations with properties of individual galaxies.

\section{Acknowledgments}
We thank the anonymous referee for helpful comments and suggestions on the paper. CLD acknowledges funding from the European Research Council for the FP7 ERC consolidator grant project ICYBOB, grant number 818940.

\bibliographystyle{mn2e}
\bibliography{Dobbs}
\bsp
\label{lastpage}
\end{document}